\documentstyle[12pt,twoside]{article}
\def\greaterthansquiggle{\raise.3ex\hbox{$>$\kern-.75em\lower1ex\hbox{$\sim$}}}
\def\lessthansquiggle{\raise.3ex\hbox{$<$\kern-.75em\lower1ex\hbox{$\sim$}}}
\newcommand{\beq}{\begin{equation}}
\newcommand{\eeq}{\end{equation}}
\newcommand{\beqa}{\begin{eqnarray}}
\newcommand{\eeqa}{\end{eqnarray}}
\newcommand{\beqan}{\begin{eqnarray*}}
\newcommand{\eeqan}{\end{eqnarray*}}
\newcommand{\ba}{\begin{array}}
\newcommand{\ea}{\end{array}}

\newcommand{\A}{{\cal A}}

\newcommand{\M}{{\cal M}}

\newcommand{\T}{{\cal T}}

\def\nz{\ifmmode {I\hskip -3pt N} \else {\hbox {$I\hskip -3pt N$}}\fi}
\def\zz{\ifmmode {Z\hskip -4.8pt Z} \else
       {\hbox {$Z\hskip -4.8pt Z$}}\fi}
\def\qz{\ifmmode {Q\hskip -5.0pt\vrule height6.0pt depth 0pt
       \hskip 6pt} \else {\hbox
       {$Q\hskip -5.0pt\vrule height6.0pt depth 0pt\hskip 6pt$}}\fi}
\def\rz{\ifmmode {I\hskip -3pt R} \else {\hbox {$I\hskip -3pt R$}}\fi}
\def\cz{\ifmmode {C\hskip -4.8pt\vrule height5.8pt\hskip 6.3pt} \else
       {\hbox {$C\hskip -4.8pt\vrule height5.8pt\hskip 6.3pt$}}\fi}
\newtheorem{theorem}{Theorem}
\newtheorem{definition}{Definition}
\newtheorem{lemma}{Lemma}

\def\au{{\setbox0=\hbox{\lower1.36775ex%
\hbox{''}\kern-.05em}\dp0=.36775ex\hskip0pt\box0}}
\def\ao{{}\kern-.10em\hbox{``}}

\normalsize
\begin{document}
\bibliographystyle{plain}

\begin{titlepage}
\begin{flushright}
\today
\end{flushright}
\vspace*{2.2cm}
\begin{center}
{\Large \bf  Interacting Fermi systems in the tracial state }\\[30pt]

Heide Narnhofer  $^\ast $\\ [10pt] {\small\it}
Fakult\"at f\"ur Physik \\ Universit\"at Wien\\
%Boltzmanngasse 5, A-1090 Wien \\

\vfill \vspace{0.4cm}

\begin{abstract}We argue that for Fermi systems on lattices or the continuum with interaction invariant under a kind of Galilei transformation the time evolution is either weakly asymptotically abelian or at least $\eta -$abelian
in the tracial state but not norm asymptotically abelian.

\\
\hspace{1.9cm}

\end{abstract}
\end{center}

\vfill {\footnotesize}

$^\ast$ {E--mail address: heide.narnhofer@
univie.ac.at}
\end{titlepage}
\section{Introduction}
In order to understand the convergence to equilibrium in the Eigenstate-Thermalization-Hypothesis-approach (ETH) \cite{MS}, \cite{JMD} local operators are expressed by eigenstates of the hamiltonian and the resulting spectrum given as difference of the eigenvalues of the hamiltonian determines the time dependence of these operators. However these eigenstates exist only for finite systems. We have to turn to the thermodynamic limit of infinite systems. This limit is best described by the $C^*-$algebra of quasilocal operators and we will concentrate on fermions with creation and annihilation operators on the lattice or on the continuum. Under appropriate assumptions the time evolution is given as a continuous automorphism group \cite{BR} or \cite{NT}. This again allows to define the Arveson spectrum \cite{Arv} of a quasilocal operator in correspondence to the spectrum for a finite system. But now this spectrum is only defined as a set, that in general is infinite and cannot be used to control convergence properties or relations between different operators. This changes if we consider time invariant states. In these states the time evolution is given by unitary operators and their spectral decomposition is inherited by the Arveson-spectrum. However the spectral decomposition depends on the invariant state. But it allows to determine some behaviour on the level of the $C^*$-algebra that may be relevant with respect to the existence of other invariant states.

As first result we will exclude for a large class of interactions the existence of operators periodic under time evolution and also the possibility that local operators remain local for all times. This can be compared to the Lieb-Robinson bound \cite{LR}  that shows that in general we can expect spreading with some finite velocity but cannot guarantee that spreading really happens for all operators. With interaction between different lattice-points operators really get delocalized. It remains to characterize this delocalisation in more detail. For quasifree evolution it is well known that the result is much stronger: creation and annihilation operators anticommute asymptotically and therefore time translated even products of creation and annihilation operators commute asymptotically in time with local operators, therefore escape to infinity.  But already for the XY-model (though not a Fermi-system but closely related to) it was shown \cite{A} that the time evolution is not norm-asymptotically abelian for the total algebra, though this holds for the even subalgebra. Nevertheless some local operators neither commute nor anticommute asymptotically, therefore  these time-translated operators consist of a local part combined with a delocalized part. Nevertheless the expectation-values of their products factorize in the limit for time-invariant states with GNS-representation that is cyclic and separating. The system is weakly asymptotically abelian.

If we examine the existing models for which it is known that only a single time invariant state exists \cite{NTS}, \cite{NT3}, \cite {BBN}, \cite{BKS}, \cite{NTBKS} in all these models neither norm asymptotic abelianess nor strong asymptotic abelianess is satisfied. Operators are delocalized but they still do not commute with some local operators. All these models do not have a physical interpretation. But they serve as a hint, that a kind of algebraic entanglement created in the course of time between distances far apart might be responsible for the approach to equilibrium as a stronger fact than approach to just some invariant state and might reduce the existence of invariant states other than equilibrium-states.

In order to study whether interacting systems are not strongly asymptotically abelian but keep non trivial commutations with local operators we will study the behaviour in the tracial state. Here the quasilocal algebra can be extended to a doubled quasilocal algebra. For quasifree evolutions we have a stronger symmetry relation that is responsible for norm asymptotic abelianess, but its violation by interaction enables to keep algebraic entanglement with local operators  in the course of time.

In this note we first define in section 1 the systems we want to consider: Fermi-lattice systems and Galilei invariant continuous systems, i. e. systems for which time evolution is defined as approximately local automorphisms. In section 2 we discuss the possibilities how time evolution can be localized or delocalized. In section 3 we work in the representation given by the tracial state and transfer the result on continuous systems \cite{NT2} that the systems gets delocalized in the course of time to Fermi-lattice systems. In section 4 we show that in the course of time operators do not get completely delocalized if the system is interacting.

Finally we discuss why the failing of strong asymptotic abelianess might be of relevance for thermodynamical behaviour.
For odd operators that anticommute it is well know that their expectation value has to vanish in invariant states. Now we increased the class of operators that do not asymptotically commute on the algebraic level. We give hints how these algebraic correlations can give additional restrictions on the invariant states based on the arguments used for anticommutativity in comparison to the consequences of randomness in \cite{BBN}.

\section {The Fermi-algebra }
The Fermi algebra is built by creation and annihilation operators $a^*(f), a(g)$ satisfying
\beq a(f)a(g)+a(g)a(f)=0, \quad a^*(f)a(g)+a(g)a^*(f)=\langle g|f\rangle \eeq
where $f$ is a square-integrable function either on the lattice or on the continuum. The even algebra $\A_e$ is built by even polynomials in creation and annihilation operators. We pick some $f_0$ which defines the unitary \beq U_0 =a(f_0)+a^*(f_0) \quad U_0^2=1 \eeq
This unitary implements the automorphism $\alpha _0(A) =U_0AU_0$ on the algebra and also a non-inner automorphism on the even algebra. It evolves in time. We are interested in its limit for large times. Anticommutativity on the odd algebra implements commutativity on the even algebra and holds, if the automorphism implemented by $\tau _t U_0$ becomes trivial.

We will mainly concentrate in our proofs on Fermi algebras on the lattice. We consider time evolutions that are implemented by
\beq H _{\Lambda }=\sum _{x,y} a^*_xa_y f(x-y)+\sum _{xyz}a^*_x a^*_y a_z a_{-x-y-z}v(x,y,z) \quad x,y,z \in \Lambda\eeq
with taking the limit $\Lambda $ becoming the total lattice, which defines a  time-automorphism-group that is invariant under space translations. \cite{BR}.   We are inspired on results for continuous Fermi-systems with Galilei-invariance. Here  the time evolution is given by the hamiltonian in dimension $\nu$
\beq
\label {FG}H= \frac{1}{2m}\int dx^{\nu}
\nabla  a^*(x)\nabla a(x)+\int d^{\nu}pd^{\nu}p'd^{\nu}qd^{\nu}q' a^*_{pq}a^*_{p'q'}w(p-p')v(q-q')a_{p'q'} a_{pq}=K+V.\eeq
 Here $a_{pq}=a(W(p,q)f)$ is an annihilation operator smeared with an $f$ that is translated by the one-particle Weyl operators $W(p,q)=e^{i(qP+pX)}.$ As concrete example we take as in \cite{NT} $f$ as a Gauss function so that with appropriate normalization in $\nu $ dimensions  with the notation
$$[a^*(x),a(y)]_+=\delta ^{\nu }(x-y)$$
$$a(f(x))=a(f)=\int d^{\nu} x f(x)a(x)$$
$$a_{pq}=\pi^{-\frac{\nu}{4}}\int d^{\nu }xe^{-\frac{(q-x)^2}{2}+ipx}a(x)=a(|p,q\rangle )$$
where $|p,q\rangle $ are coherent states.
The cutoff in the interaction is introduced to guarantee that the time-evolution exists as continuous automorphism-group. This is shown as for the lattice-system on the basis of perturbation-theory.

\section{Localization and delocalization}
Classical dynamical systems are given by a probability space equipped with a finite probability measure $\mu$ and a continuous measurable map $T(t)$ for which the measure is invariant $\mu \circ T_t=\mu.$ The system is ergodic if the average over $f(T_tx)$ is $\mu$-almost constant. This can be improved by demanding that the average is replaced by the limit $t\rightarrow \pm \infty.$
The quantum analog is a quasilocal $C^*$-algebra, here the Fermi-algebra built by creation and annihilation operators $a(f),a^*(g)$
satisfying \beq a(f)a(g)=-a(g)a(f),\quad a(f)a^*(g)+a^*(g)a(f)=\langle f|g\rangle \eeq
where $f(x)$ is a function either over $R^{\nu}$ or over the lattice. Space-translation is given by $\sigma _x a(f(y))=a(f(x+y)).$ Local subalgebras $\A_{\Lambda}$ are built by creation and annihilation operators with support of $f$ in $\Lambda.$ Time evolution is given for lattices by \beq  \frac{d}{dt}\tau _t A= i[\sum \sigma_x H_{\Lambda },A]\eeq\
with $H_{\Lambda }$ localized in $\Lambda $. The existence of the time evolution as continuous automorphism-group on the algebra is guaranteed \cite{BR} and this can be generalized for $H$ sufficiently approximated by local operators and also for continuous systems with interactions where the point interaction is slightly smeared out \cite {NT}.

\begin{definition} The time evolution is strictly localized if for any local operator $ A \in \A_{\Lambda }$ there exists a $\bar{\Lambda }$ such that $\tau _tA\in \A_{\bar{\Lambda}} \forall t$.
This can be characterized by \beq [\tau _t A, B]=0 \forall t\in R, \forall B\in \A_{\bar{\Lambda } ^C}, B \quad even\eeq
.\end{definition}
This holds evidently if there is no interaction between different lattice points.
\begin{definition} The time evolution is delocalized, if it is not strictly localized.\end{definition}
This can be specified:
\begin{definition} The time evolution is completely delocalized, if for every localized A and every $ t_0\in R$ there exists a $\Lambda $ such that \beq [\tau (t)A, B]=0 \quad \forall t>t_0, \forall B\in \A_{\Lambda } \eeq\end{definition}

A slightly weaker form is
\begin{definition}The evolution is  norm asymptotically abelian on the even subalgebra , if
\beq norm-\lim [\tau _t A,B]=0 \quad A \quad even.\eeq \end{definition}
This is shown for quasifree time evolutions $\tau _ta(f)=a(e^{iht}f)$ where $h$ has absolutely continuous spectrum that is nowhere constant \cite{N70}.
Similarily we can define strong and weak asymptotical abelianess by replacing the norm limit by the strong limit in a representation of the algebra or by the weak limit. As it will turn out, which kind of limit holds will depend on the representation of the algebra, whereas norm limit is independent of the representation.

The fact that the group $R$ is ameanable guarantees that invariant states exist. For the considered time evolution also equilibrium states satisfying the KMS condition exist obtained by limiting considerations \cite{BR}. Asymptotic abelianess also in its weaker form can be used to obtain extremal invariant states \cite{D}. It also guarantees that local perturbations of time invariant states return to the unperturbed state. Further it can be used under the additional assumption of dynamical stability \cite{HKTP} or passivity \cite{PW} to prove that the extremal invariant state is an equilibrium state. However the fact that quasifree states that are space-translation invariant are also invariant under all quasifree time translations that commute with space translations shows that norm asymptotic abelianess cannot explain approach to equilibrium without further restrictions on the initial state.

\section{Delocalization for interacting systems}
For continuous Fermi systems with Galilei invariant interaction it is shown \cite{NT2} that they are weakly asymptotically abelian in the tracial state. The proof is based on the fact that acting on the system with the Galilei automorphism $\gamma (g)a(f(x))=a(e^{igx}f(x))$ combined with the Galilei invariant time evolution space translation is added to the free evolution $\gamma (g)\tau _0(t)\gamma (-g)a((f)= \sigma (gt) \tau _0(t) a(f)$. The interaction was adjusted to keep Galilei invariance, which connects space, boost, gauge and time transformations
 \beq\sigma _xa(f(y))=a(f(x+y))\quad \gamma _b a(f(y))=a(e^{iby}f(y))\quad
\nu _{\alpha }a(f)=e^{i\alpha }a(f) \quad
 \tau _t,\eeq
such that
$$ \sigma _x \circ \nu _{\alpha } =\nu _{\alpha } \circ \sigma _x, \quad \gamma _b \circ \nu _{\alpha } =\nu _{\alpha} \circ \gamma _b, \quad \gamma _b \circ \sigma _x=\sigma _x \circ \gamma _b \circ \nu _{-bx}$$
$$ \tau _t \circ \nu _{\alpha }=\nu _{\alpha } \circ \tau _t, \quad \tau _t \circ \sigma _x =\sigma _x \circ \tau _t \quad \tau _t \circ \gamma _b =\gamma _b \circ \tau _t \circ \sigma _{bt} \circ \nu _{-b^2t/2}.$$
The advantage of the above model is the fact that  the time evolution is related to the space translation by Galilei invariance so that it can inherit weak asymptotical abelianess \cite{NT2},\cite{NT3}: small perturbation by the Galilei automorphism adds a contribution of space translations and the possible separation of space-translation and time-translation combined with a smearing of the Galilei transformation implies that the decay of the spacial correlation functions is inherited by the time correlation functions. This tells us especially, that no local operator remains strictly local in the course of time. Notice that only one-time correlations could be controlled, multi-time correlations remained as an open problem.

With an appropriate replacement of the Galilei-automorphism we can prove a slightly weaker result for lattice systems:
\begin{theorem} Let the time evolution be implemented by terms of the form $\sum a^*_{x_1}..a^*_{x_k}a_{y_1}..a_{y_k}\delta (x_1 +..x_k -y_1 -..y_k). $ Consider the automorphism $\gamma (g) a_x=e^{igx}a_x$. Then the time evolution in the tracial state has apart from the GNS vector $|\Omega \rangle $ a continuous spectrum. \end{theorem}
Proof: By assumption on the time evolution
\beq \sigma (x) \gamma (g) \tau (t) \gamma (-g)\sigma (-x)=\gamma (g) \tau (t) \gamma (-g) =\tau _g (t)\eeq
The tracial state is invariant under all automorphisms that can be locally approximated. Therefore these automorphisms are implemented by unitaries that let the GNS vector $|\Omega \rangle $ invariant. Assume $\tau (t)A|\Omega \rangle =e^{iEt}A|\Omega \rangle$ for some operator $A$ that can be approximated arbitrary well by a local operator. Choose $g$ such that $||\gamma (g)A -A|\Omega \rangle||< \epsilon .$ Then also $$||\tau (t) (\gamma(g)A-A)||=||\gamma (-g) \tau (t) (\gamma (g)A-A)|\Omega \rangle || =||\tau _g(t)A-e^{iEt}\gamma (-g)A|\Omega \rangle ||<2\epsilon.$$

By assumption on the time evolution also $\tau _g(t)$ commutes with space translation. Therefore with $ \sigma _n \gamma (g) \sigma _{-n} = \gamma _n(g)$ also $\gamma _n(g) \tau (t) \gamma _n(-g)= \tau _g(t)$ and thus
$$||(e^{iHt}-e^{iEt})e^{igG_n}A|\Omega \rangle ||\leq \epsilon \quad \forall n,t, g\leq g_0$$

With increasing $n$ $\gamma _n(g)$ rotates locally with increasing velocity. Therefore varying over $t$ and $n$ it follows that $A$ does not only approximate an eigenvector for $H$ with eigenvalue $E$ but also an eigenvector for $G$ with eigenvalue $0.$ Such an eigenvector corresponds to a quasilocal operator of the commutative subalgebra built by $a^*_x a_x $ . But according to our choice of the hamiltonian this excludes that they are eigenoperators of the time evolution.

The fact, that there are no eigenvectors and therefore also no eigenoperators for the time evolution  implies
\begin{theorem}For interacting systems as in Theorem (1) there are no local operators that remain strictly local in the course of time.\end{theorem}
Proof: For a unitary-group with only continuous spectrum every subspace that is invariant under the evolution is infinite dimensional. Local operators create a finite dimensional subspace. Therefore no vector respectively operator can stay in a finite dimensional subspace respectively local algebra.

Whereas for the continuous system it was shown that time correlations $\omega (A\tau(t)B)-\omega (A)\omega (B)$ vanish for $t\rightarrow \infty $ and therefore the system is weakly asymptotically abelian at least in the tracial state, we only obtain that the spectrum of the unitary implementing the time automorphism is apart from the GNS vector continuous. We did not succeed to prove that it is absolutely continuous so that we could apply Riemann-Lebesgue. A further argument on the control of the continuous spectrum would be necessary. It remains so far that the invariant mean of local operators $\eta \tau (t)A = \omega (A)$ and commutes with all local operators.

\section{The tracial state as the Vacuum state over an extended algebra}

In \cite{J} it was shown that applying the results of scattering theory in the vacuum state the time automorphism also for interacting systems is though not in norm so nevertheless strongly asymptotically abelian. Scattering theory does not exist in temperature states \cite{RNT}, and in general we do not have a replacement to control long time behaviour. This is different for the tracial state, that is invariant under all approximately local  automorphisms and allows a concrete construction of the GNS-representation that will enable us to perform estimates also for interacting systems.

Let $V\in\A$ with $V=V^*$ implement an automorphism on $\A .$ In the GNS-representation of the tracial state let \beq V|\Omega \rangle =V'|\Omega \rangle \quad V' \in \A'\eeq
\begin{theorem} Under the assumption that $st-\lim _{t\rightarrow \pm \infty }\tau (t)V A \tau (t)V =A$ it follows that $st-\lim _{t\rightarrow \pm \infty }\tau (t)V\tau (t)V'=1$ \end{theorem}
Proof: \beq st-\lim_{t\rightarrow \pm \infty }\tau (t)VA\tau (t)V|\Omega \rangle =A|\Omega \rangle =\eeq $$=st-\lim _{t\rightarrow \pm \infty }\tau (t)VA\tau (t)V'|\Omega \rangle =st-\lim _{t \rightarrow \pm \infty }\tau (t)V\tau (t)V'A|\Omega \rangle \quad \forall A\in \A $$
where $A|\Omega \rangle $ is dense in the Hilbertspace.  $V$ itself converges weakly to $0$. But in the tracial state it is not necessary to control its action as automorphism. By combination with the commutant we have only to control the convergence of a well defined unitary operator to $1.$

If we replace $V$ by an operator that is assumed to be asymptotically anticommutative, then with \beq WA_e|\Omega \rangle =A_e|\Omega \rangle \quad WA_o|\omega \rangle =-A_o|\Omega \rangle \eeq for even and odd operators we have to replace $st-\lim _{t\rightarrow \pm \infty }\tau (t)V\tau (t)V'=W.$

We turn to our system of interest and specify $V $ and $V':$
\begin{theorem} With the observable algebra $\A(a)$ built by creation and annihilation operators $a(f),a^*(f) $ and the algebra $\A(b)$ built by $b(f)=WJa(\bar{f})J $, J the modular conjugation, the GNS representation of the tracial state coincides with the vacuum state over an extended Fermi-algebra built by creation and annihilation operator $A(f),A^*(f),B(f), B^*(f)$ connected with the initial algebra by a Bogoliubov transformation
\beq a(f)=\frac{1}{\sqrt2}(A(f)+ B^*(\bar{f})), \eeq
With
\beq b(f)=\frac{1}{\sqrt2}(A(f) -B^*(\bar{f}))\eeq
\beq [a(f),b(g)]_+=0,[a(f), b^*(g)]_+=0,[b(f),b(g)]_+=0,[b(f)^*,b(g)]_+=\langle g|f\rangle \eeq
The even algebra $\A(b)_e$ commutes with $\A(a).$
Let $|\Omega \rangle $ be the vacuum vector for $\A(A,B)$ and $H_e$ respectively $H_o$ be the subspaces built from $|\Omega \rangle $ by even respectively odd polynomials in $A,B,A^*,B^*$ and let \beq W |\Psi _e\rangle =|\Psi _e \rangle \quad W|\Psi _o \rangle=-|\Psi _o\rangle, \quad |\Psi _{e,o}\rangle \in H_{e,o} \eeq
then $Wb(f), b^*(g)W$ commute with $\A (a)$ and create the commutant $\A(a)'$. \end{theorem}
Proof: Both the vacuum state over $\A(A,B)$ and the tracial state over $\A(a)$ are determined by the two point function \beq 2\omega (a(f)a^*(g))=\langle g|f\rangle =\frac{1}{2}\omega (A(f)+B^*(\bar{f})(A^*(g)+B(g)))\eeq
Further with $J$ the modular conjugation
\beq a(f)|\Omega \rangle =Ja(\bar{f})|\Omega \rangle =\frac{1}{\sqrt2} B^*(\bar{f})|\Omega \rangle =Wb(f)|\Omega \rangle, \quad Ja(\bar{f})J=Wb(f). \eeq
The time evolution is implemented by an operator satisfying $H=-JHJ, H|\Omega \rangle =0$ . Acting on $\A (a)$ it is implemented by a sequence of local operators $h_{\Lambda }$ that we can express term by term by $A(f),B(f),A^*(g)B^*(g)$ First we concentrate on  quadratic terms that contribute to $h_{\Lambda }$:
\beq 2a^*(f)a(g)=A^*(f)A(g)+B(\bar{f})A(g)+A^*(f)B^*(\bar{g})+B(\bar{f})B^*(\bar{g})\eeq
These terms do not annihilate the vacuum. However the effect on $\A(a)$ remains unchanged if we add counterparts from the commutant, including c-numbers:
\beq b^*(f)b(g)=A^*(f)A(g)-B(\bar{f})A(g)-A^*(f)B^*(\bar{g})+B(\bar{f})B^*(\bar{g})\eeq
Therefore we choose as contributions to the hamiltonian
\beq a^*(f)a(g)+b^*(f)b(g)-\langle g|f\rangle =A^*(f)A(g)-B^*(\bar{g})B(\bar{g})\eeq
Notice that the necessary c-number renormalization tends to infinity in agreement with the fact that the time evolution is not an inner automorphism of $\A(a).$
We observe that a quasifree time evolution on $\A(a)$ becomes a quasifree time evolution also on $\A(A,B)$ and decouples between $\A(A)$ and $\A(B).$ Notice that different to the vacuum state the hamiltonian implementing the time-evolution for the extended algebra is not bounded from below. It is by construction invariant under the modular conjugation combined with the reflection $\gamma A(f)=B(f)$, which coincides with time reflection individually in $\A(a)$ and $\A(b)$. Therefore time limits in different time-directions can coincide. On the basis of the $C^*$ algebra we have convergence in norm for the automorphisms implemented by $a_0+a^*_0$, on the basis of the corresponding unitary given in Theorem (3) we can write it as \beq U=VV'W=2P-1\eeq where expressed in the Bogoliubov-transformed creation and annihilation-operators  \beq P=A_0A_0^*+B_0B_0^* \eeq and we have strong convergence of $P$ to unity in both time-directions and similarily strong convergence to $0$ for $1-P.$ This holds for quasifree time evolution and we have to examine what changes for interacting systems.
In more generality we observe:
\begin{lemma}Let $U $ be a unitary with $U=U^*.$ Let  $st-\lim _{t\rightarrow \pm \infty }\tau(t)U=1$. Then with $U=2P-1$ $st-\lim _{t\rightarrow \pm \infty }\tau _t P=1$, whereas with $Q=1-P$ $st-\lim _{t\rightarrow \pm \infty }\tau _t Q=0.$  \end{lemma}
Proof: In the strong limit eigenvalues can disappear but cannot be created. Therefore $U$ must contain an infinite dimensional projection, that in both time directions converge to one and has therefore to be invariant under time reverse.
\begin{lemma} Assume that $\tau _tA =e^{iHt}Ae^{-iHt}$ is strongly asymptotically abelian on the even algebra. Then it follows that $st-\lim_{t\rightarrow \pm \infty} [H,P]=0.$ \end{lemma}
Proof: Under the assumption of strong asymptotically abelianess according to (24) and the lemma
$$st \lim _{t\rightarrow \infty }e^{iHt}(P-e^{i\epsilon H}Pe^{-i\epsilon H})e^{-iHt}=0 \forall \epsilon$$
Taking the limit $\epsilon \rightarrow 0$ reduces to convergence of the time derivative.

We imply these observations to time evolution without and with interaction. Since we know, that quasifree automorphisms are norm asymptotically antiabelian, in the combination $U=(a_0+a^*_0)(b_0-b^*_0)$ the time-evolution of this unitary satisfies the requirements in the lemma and converges in the indicated way. Expressed in the Bogoliubov transformed algebras convergence happens separately for the gauge invariant parts of $\A(A)$ and $\A(B)$. However interaction mixes between these gauge invariant parts. We concentrate on the contributions of the interaction. They are given by polynomials of forth order in creation and annihilation operators. Expressed by $\A(A,B)$ they contain contributions with four creation operators, with three creation operators with two creation operators and with one creation operator. The contribution from the commutant necessarily has to annihilate the term with four creation operators in order to satisfy $H|\Omega \rangle =0.$
This holds if we choose \beq a^*(f)a^*(g)a(g)a(f)-b^*(f)b ^*(g)b(g)b(f)\eeq and observe that it contains terms like
\beq A^*(f)B^*(f)B^*(g)B(g)\eeq
where we use that for the expression following from the contributions of $b$ the terms with four, two and zero creation operators cancel and only the terms with three and one remain, where in addition the annihliation operator has to be commuted to the right which gives an additional quadratic contribution.
We observe two facts: On one hand the time evolution on the algebra $\A(b)$ is given by the kinetic part together with an interaction similar as for the observable algebra but with opposite sign (Compare (20) and (21)). If we believe that the sign of the interaction is relevant for stability properties of the system, we might expect that the resulting instability in the commutant can have consequences also in the algebra. On the other hand the tracial state for the commutant is the limit of increasing negative temperature, for which stability arguments are turned around. Relevant is the fact, that time reflection takes place in the algebra of observables, whereas the modular conjugation acts as time reflection combined with a reflection between algebra and commutant. For quasifree time evolution we have the additional symmetry between $\A (A)$ and $\A(B)$ and thus the necessary relation between modular conjugation and time-reflection. That does not hold for interacting systems. If we express the hamiltonian by the operators $A$ and $B$, it is not the sum of operators belonging either to $\A(A)$ or to $\A(B)$. Further it is not invariant under the gauge transformation of either $\A(A)$ or $\A(B).$

The interaction contains contributions
$$(A_x-A_x^*-B_x^*+B_x)(A_y+B_y^*+A_y^*+B_y)=2(A_xB_y^*+B_x^*A_y+A_xA_x^*)$$
and corresponding terms reflecting the symmetry between $\A(A)$ and $\A(B)$ that is opposite to the symmetry for the quasifree evolution.
They contribute to the timederivative
$$ \frac{d}{dt}P=i[H,( A_xA_x^*+B_xB_x^*)])$$
Especially for strictly local interaction the time derivative remains strictly local and can be evaluated to be an operator without eigenvalue $0$. Therefore it can converge weakly to 0, but not strongly, as it would be necessary and we run into a contradiction for strong asymptotic abelianess.
  This already can be seen in the model of \cite{GGE} with $H=\sum_x \Pi_{j\in\Lambda }\sigma _z^{j+x}f(\Lambda )$ where the quasifree part of the hamiltonian is missing. Here the system is  even strictly localized. We have an abelian subalgebra that is pointwise invariant under time-evolution. The remaining operators, especially the creation and anhiliation operators for the Fermi system respectively $\sigma_x, \sigma _y$ localized at the point $0$ for the corresponding lattice system converge weakly to $0$ \cite {Ra} though they stay localized. Delocalization is produced by adding the quasefree part to the hamiltonian and this delocalization is improved by the interaction creating nontrivial commutation-relations between local regions that are far apart.

We summarize our observations in the tracial state and their consequences for the $C^*-$ algebra:
\begin{theorem}For Fermi lattice systems with interaction the time evolution is not strictly localized but also not norm asymptotically abelian. \end{theorem}
As a by-product we observe, that the kind of asymptotic abelianess can be state-dependent, in the groundstate we have strong asymptotic abelianess \cite{J}, in the tracial state at its best weak asymptotic abelianess.

\section{Consequences}
We have demonstrated that for quasifree and for interacting systems local operators do not remain local for increasing time. Comparing with a strictly local evolution where we can construct invariant states by choosing locally an invariant state depending on the position and take arbitrary products of these states, nonlocality produces constraints between far separated regions. We observe, that there are more constraints if the time evolution is not strongly asymptotic abelian:   Choosing a subalgebra $\M_{\nu }$ with dimension $\nu$ of our observable algebra we can write \beq \A(a)=\M_{\nu } \otimes \A _M\eeq
With $m\in \M_{\nu }$ we consider $(\tau _t m)_{ik}=$ as element in $\A_M$. If the time automorphism is norm asymptotially abelian then $\lim _t [(\tau _t m)_{ii}-(\tau _t m)_{jj}]=0,\lim _t (\tau _t m)_{ij}=\lim _t (\tau _t m)_{ji}=0.$
If they do not commute then $||(\tau _t m)_{ij}|| > 0$ for arbitrary large $t$.  An invariant state is given by functionals $ \omega _{ik} $ over $\A_M$ with $\omega (m_{ij})=\omega _{ij}.$
$ (\tau _t m)_{ij}$ varies with $t$ and time invariance implies constraints for $ \omega _{ik} $ If the algebraic correlations that exist for interacting systems and are not trivial but are random which was assumed for all operators in the case of \cite {BBN} then it follows that $\omega _{ij}=0=\omega (m_{ij})$ and we obtain a condition on the local level. Similarily correlations between annihilation operators $a(f)$ give restrictions, for example gauge invariance of the state.  In general it seems plausible  that with increasing number of algebraic correlations also the possibility for an invariant state is reduced.

Galilei invariance allowed that spacial clustering is inherited by time clustering. However the proof fails for multiclustering, and in fact for norm asymptotically abelian time evolution \beq \lim _{t\rightarrow \infty }\omega (AB_tCD_t) = \omega (AC)\omega (BD)\omega (AB_tCD_t) = \omega (AC)\omega (BD) \eeq
whereas for a time evolution that is weakly asymptotically abelian or $\eta-$abelian \beq \lim _{t\rightarrow \infty }|\omega (AB_tA^*B^*_t)| < \omega (AA^*)\omega (BB^*). \eeq
Let us assume that  states extremely invariant under space translation cluster uniformly and this clustering is inherited by time translation
\beq \lim _{t\rightarrow \infty } P_{\Lambda_2 ^C}e^{iHt}P_{\Lambda _1}=|\Omega \rangle \langle \Omega | \eeq
where $\Lambda_2$ containing $\Lambda_1$ are local regions,$\Lambda^C$ the complement of $\Lambda$ and $P_{\Lambda }$ the projection operator onto $\A_{\Lambda }|\Omega \rangle.$
Under this assumption together with norm asymptotic abelianess
\beq \lim_{t_j -t_i \rightarrow \infty } \omega (\tau _{t_1} A_1\tau _{t_2} A_2...\tau _{t_k} A_k)=\omega (A_j)\omega_{i\neq j} (\Pi \tau _{t_i} A_i)\eeq
independently of the ordering and the size of $t_i$ and we can control multiclustering. This makes it possible to control local perturbations
\beq \langle \Omega |Ae^{it(H+\lambda V)}Be^{-it(H+\lambda V)}A^*|\omega \rangle .\eeq By expanding in the coupling constant and taking the limit $t\rightarrow \infty $ we can obtain an invariant state. In KMS-states where the time automorphism coincides with the modular automorphism we can construct this invariant state by the time ordered expression
\beq |\Omega _V\rangle = \T \int _0^{\beta/2} d\gamma e^{\tau_{i\gamma }V}|\Omega  \rangle \eeq
where now the integration runs over a finite region and does not refer to asymptotic behaviour, therefore can also be applied for systems that are just weakly asymptotically abelian.
An other possiblity to construct the state invariant under the perturbation but staying in the follium is looking for the vector in the natural cone satisfying
\beq (H+\lambda (V-JVJ))|\Omega _V\rangle =0=|\Omega +\sum \lambda ^n B_n\rangle \eeq which is solved by using the modular operator $M$ that we now may assume to be different from $H$ but by invariance commutes with $H$. Expanding in $\lambda $ gives
\beq B_1|\Omega \rangle =-\frac{1}{H}(V-JVJ)|\Omega \rangle =-\frac{1}{H}(1-e^{-M/2}) V|\Omega \rangle \eeq
\beq B_n|\Omega \rangle =\frac{1}{H}B_{n-1}|\Omega \rangle .\eeq
Considering
\beq \lim _{\epsilon \rightarrow 0} \frac{-i}{H+i\epsilon} =\lim _{\epsilon \rightarrow 0}\int_0^{\infty } dt e^{it(H+i\epsilon)} \eeq
we are dealing with an unbounded operator $\frac{1}{H}$, but from clustering in time it follows that $(V-e^{-M/2}V)|\Omega \rangle $ belongs to its domain. However for $B_n|\Omega \rangle$ we would need higher order of time correlation functions that for weakly asymptotically abelian or $\eta-$abelian systems are not available.

Therefore we have to be aware that differently as for KMS-states local perturbations of the dynamics can lead to states that are not normal with respect to the initial state, i.e. that already the local perturbations can create infinite energy in the course of time. We observe different behaviour with respect to the degree of abelianess, but also the special role of KMS-states. Extending the local perturbation to global perturbation as it is possible for KMS-states without violating the KMS-property with respect to the new dynamics is even more problematic. All these are indications of the sensibility of weakly asymptotically systems and increasing restrictions for invariant states.

\section{Conclusion}
We have concentrated on Fermi-systems on the lattice. Our considerations can with appropriate modifications be generalized to lattice systems and Galilei-invariant lattice systems. For these systems the tracial state exists. In  the corresponding GNS-representation every approximately local automorphism can be represented by unitaries. The relevant automorphisms are space translation, time translation with and without interaction, Galilei-transformation respectively its replacement for lattice-systems and gauge-transformation. The relation between time evolution and Galilei transformation was used to prove, that no local operator is an eigenoperator of the time evolution and consequently becomes delocalized in the course of time.
Next we used gauge-translation and space translation to construct a local structure also on the commutant and extended this local structure to all operators in the representation, including the operators of the commutant. The modular conjugation which is an antilinear automorphism of the total algebra respects the local structure. The relation between modular conjugation and time reflection allows strong abelianess for quasifree evolution, but for interaction that react differently on the antilinearity algebraic correlations between local operators and local operators shifted in time remain.

Finally we search for arguments why weak asymptotic abelianess might reduce the possibility for time invariant states as suggested by examples of discrete time evolution. On one hand it offers additional areas of randomness that create restrictions on space correlations. On the other hand we refer to the demand of dynamical stability i.e. the existence of invariant states under local perturbations without leaving the follium. This holds for KMS-states, but otherwise the necessary analyticity properties get out of control for interacting systems.

We are optimistic that further analysis that concentrates on long range effects in time on the algebraic level and their effect on long range correlations in space for invariant states
can lead to deeper understanding of the importance and the power of interaction with respect to thermodynamics.

\bibliographystyle{plain}

%\tableofcontents
%\makeindex
\end{document}